\documentclass[aps,prb,twocolumn,superscriptaddress]{revtex4-1}
\usepackage[pdftex]{graphicx}
\usepackage{booktabs}
\usepackage{multirow}
\usepackage[mediumspace,mediumqspace,Grey,squaren]{SIunits}
\usepackage{color}
\usepackage{natbib}
\usepackage{amsmath,amssymb}
\usepackage{lipsum}
\usepackage[normalem]{ulem}
\usepackage{bm}
\graphicspath{ {./Figures/} }

\begin{document} 


\title{\textcolor{black}{Driving magnetization dynamics in an \\ on-demand magnonic crystal by magneto-elastic interaction}}

\author{C.L. Chang}
\thanks{Authors contributed equally to this work}
\affiliation{Zernike Institute for Advanced Materials, University of Groningen, Groningen, The Netherlands}
\author{S. Mieszczak}
\thanks{Authors contributed equally to this work}
\affiliation{Faculty of Physics, Adam Mickiewicz University in Poznan, Pozna\'{n}, Poland}
\author{M. Zelent}
\affiliation{Faculty of Physics, Adam Mickiewicz University in Poznan, Pozna\'{n}, Poland}
\author{V. Besse}
\affiliation{IMMM UMR CNRS 6283, Le Mans Universit\'{e}, Le Mans cedex, France}
\author{U. Martens}
\affiliation{Institute of Physics, University of Greifswald, Greifswald, Germany}
\author{R.R. Tamming}
\affiliation{Zernike Institute for Advanced Materials, University of Groningen, Groningen, The Netherlands}
\author{J. Janusonis}
\affiliation{Zernike Institute for Advanced Materials, University of Groningen, Groningen, The Netherlands}
\author{P. Graczyk}
\affiliation{Faculty of Physics, Adam Mickiewicz University in Poznan, Pozna\'{n}, Poland}
\author{M. M{\"u}nzenberg}
\affiliation{Institute of Physics, University of Greifswald, Greifswald, Germany}
\author{J.W K{\l}os}
\email{klos@amu.edu.pl}
\affiliation{Faculty of Physics, Adam Mickiewicz University in Poznan, Pozna\'{n}, Poland}
\affiliation{Institute of Physics, University of Greifswald, Greifswald, Germany}
\author{R.I.Tobey }
\email{raanan.tobey@gmail.com}
\affiliation{Zernike Institute for Advanced Materials, University of Groningen, Groningen, The Netherlands}
\affiliation{LUMOS, Center for Integrated Nanotechnologies, Los Alamos National Laboratory, Los Alamos, USA}


\begin{abstract}
\textcolor{black}{Using spatial light interference of ultrafast laser pulses, we generate a lateral modulation in the magnetization profile of an otherwise uniformly magnetized film, whose magnetic excitation spectrum is monitored via the coherent and resonant interaction with elastic waves.  We find an unusual dependence of the magnetoelastic coupling as the externally applied magnetic field is angle- and field-tuned relative to the wavevector of the magnetization modulation, which can be explained by the emergence of spatially inhomogeneous spin wave modes. In this regard, the spatial light interference methodology can be seen as a user-configurable, temporally-windowed, on-demand magnonic crystal, potentially of arbitrary two-dimensional shape, which allows control and selectivity of the spatial distribution of spin waves.  Calculations of spin waves using a variety of methods, demonstrated here using the Plane Wave Method and Micromagnetic Simualation, can identify the spatial distribution and associated energy scales of each excitation, which opens the door to a number of excitation methodologies beyond our chosen elastic wave excitation.    
 }
\end{abstract}
\maketitle

\section{Introduction}
The magnetic excitation spectrum of a thin (tens of nanometers), uniformly magnetized film is well-studied and understood\cite{Gurevich,Arias16}. The fundamental {\it spin wave} mode, where the magnetization precesses in phase (i.e. with the wave vector $k=0$) in the entire volume of the system, is called the Kittel mode and can be measured by experimental techniques such as ferromagnetic resonance or time resolved magneto-optical Kerr (or Faraday) effects\cite{vanKampen2002}.  One can also observe spin wave (SW) confinement (and quantization) along the film depth, whose energy depends on film thickness and pinning effects at the surfaces. These modes, called perpendicular standing SW modes (PSSW), are still laterally uniform in amplitude and phase\cite{vanKampen2002} for in-plane wave vector $k=0$.

Going beyond lateral phase homogeneity brings about the appearance of spin wave modes of finite wavevector ($k>0$). At low wavevectors, the spin wave dispersion is highly anisotropic with respect to the direction of an externally applied magnetic field due to dominating dipolar interactions\cite{Hurben1995, Kalinikos1986}. With increasing wavevector, exchange interactions become more important and the SW modes are termed isotropic exchange SW.  Depending on the particulars of the dispersion relation (determined by magnetic field orientation, and the relative strength of dipolar to exchange interactions), the SW modes can have positive, zero, or negative group velocity. \textcolor{black}{The SW dispersion, in both dipolar and exchange regimes, can be found by optical means using Brillouin spectroscopy\cite{Demokritov01}, while time-resolved magneto-optical imaging based on Faraday effect can  also be used to determine the dispersion of SW \cite{Hashimoto}.  This latter technique is limited to purely dipolar SWs due to spatial resolution limits associated with the particular probing wavelength that is used. }

Structuring the lateral magnetic landscape further modifies the SW spectrum, while opening opportunities for spin wave localization and control and manipulation. This is the scientific discipline of $\it{magnonics}$, where artificially engineered and spatially patterned magnetic materials such as arrays of magnetic dots\cite{Tacchi11}, holes in magnetic films (antidots)\cite{Mandal12,Zelent17}, magnetic stripes\cite{Mruczkiewicz2017,Sadovnikov16,Topp2010}, and bicomponent arrays\cite{Krawczyk13}, as well as more complex spatial patterns such as magnonic quasicrystals\cite{Rychly2016}, as well as the inherent domain structure of multilayers of high perpendicular anisotropy\cite{Banerjee2017} can be used to manipulate and localize SW dynamics. The periodic magnetization profile forms a so-called $\it{magnonic~crystal}$.

For magnonic crystals (MCs) operating in dipolar regime, the dispersion relation can be tuned by the change of the direction of external field with respect to the periodically patterned structure\cite{Gallardo2014,Banerjee17}
This effect results from the  presence of static and dynamic (de)magnetizing fields and is observed even for standing ($k=0$) SW modes. The additional feature of MCs is a band structure of SW dispersion which is manifested by the presence of multiple modes for the same value of wave vector $k$ (when $k$ is reduced into the first Brillouin zone). Therefore at $k=0$ there exists a sequence of modes for which the spatial distribution of amplitude and phase within individual unit cell of MC is repeated periodically in the whole structure.

\begin{figure}[h]
    \centering     
    \includegraphics[width=.8\columnwidth]{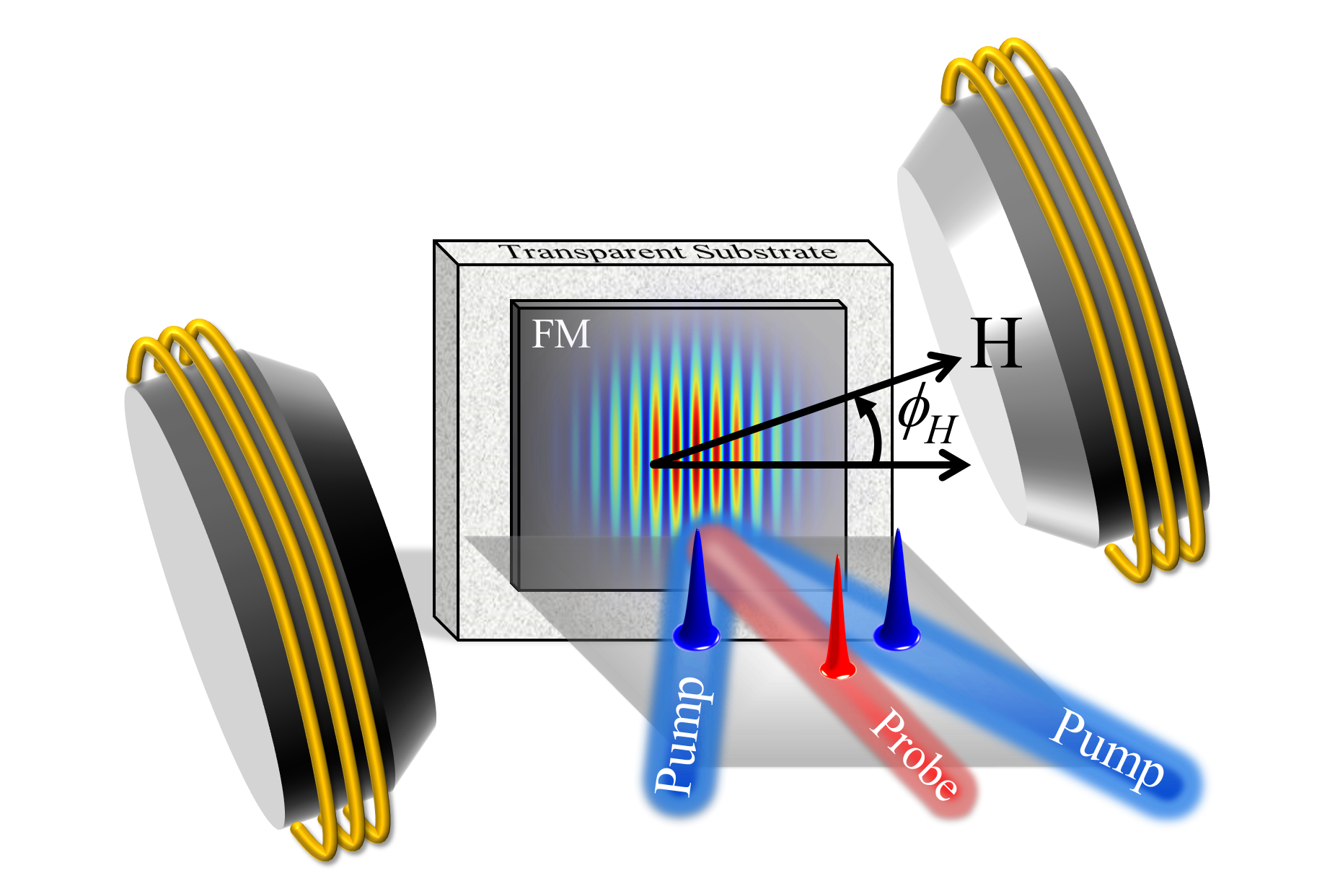}
    \caption{We excite the ferromagnetic thin film with two interfering ultrashort pulses, which simultaneously generates the surface propagating acoustic waves and laterally suppresses the magnetization profile to form the magnonic crystal. A magnetic field $H$ can be fully rotated around the sample normal and the angle $\phi_H$ denotes its direction relative to the periodicity of magnonic crystal.}
    \label{fig:experiment}
\end{figure}

Shining light onto the surface of a magnetic material can modify the magnetization landscape due to heating, thereby opening the possibility of using spatial patterning of light to induce magnonic behaviour. In most ultrafast optical experiments on magnetic materials, a large aperture optical beam is used to excite the sample surface resulting in uniform suppression of the magnetization profile, impulsively modifying the effective field landscape, and generating the laterally homogeneous free precession of a Kittel-like mode, as well as PSSW excitation\cite{vanKampen2002}. Attempts to laterally shape the excitation pattern have had success as evidenced in recent papers by Busse et. al\cite{Busse2015} and Vogel et.al.\cite{Vogel2015} utilizing continuous and pulsed laser sources respectively. In the context of magnonics, spatially patterned optical excitation offers user-defined, re-programmable arrangements of magnetic properties with the potential for unparalleled control of spin wave generation and propagation\cite{Lenk2011}. \textcolor{black}{These latter two aspects directly relate to the possibility of spin-wave signal processing\cite{Casba,Klingler}, combining two required elements, namely a spatially extended, or distributed, coherent source of SWs and the possibility to manipulate their respective phases.  Both of these requirements can be fulfilled, in principle, by our optical interference methodology}


In this report we provide a unique view of the effects of  \textit{a spatially-periodic optical excitation of a uniform magnetic film}, the emergence of a magnonic crystal, and finally, the elastic excitation of spatially distributed SW modes \cite{Graczyk2017,Graczyk2017b}.  We extend the interpretation and understanding of our previous results\cite{Janusonis2015, Janusonis2016, Janusonis2016_1} by detailing the precessional response as a function of angle of the applied magnetic field relative to the MC wavevector. In doing so, we augment our previous identification of elastically driven ferromagnetic resonance to include localized spin wave modes which exist on the magnetically modulated magnetization background. \textcolor{black}{The periodic and inhomogeneous pattern of spin wave eigenmodes allows also to change magneto-elastic interaction. We show that this non-uniformity alters the anisotropy of \textit{magneto-elastic coupling} observed in homogeneous magnetic film\cite{Dreher11,Dreher}.This additional anisotropy of magneto-elastic interaction can be explained only if the spatial modulation in the SW profiles is accounted for, in effect, an optically induced magnonic crystal.}

The present paper contains experimental and modeling sections. In the first section, we perform the transient grating (TG) experiments on two materials, Ni and CoFeB thin films (40nm films on transparent substrates such as glass or MgO), which show markedly different responses as a function of angle of the applied field for a fixed acoustic wavelength of $1.1\mu$ m. In the Ni films (low Curie temperature, $T_C$, low saturated magnetization, $M_S$) an unexpected and previously unwitnessed (binary) phase shift of $\pi$ is evident as the magnetic field angle is scanned from zero to 90 degrees (relative to the TG wavevector) [see Fig. \ref{fig:experiment}].  Accompanying this evolution in phase is a strong suppression in precessional amplitude in the intermediate region; indicative of interference between two (or more) distinct modes of precession in two different angular regimes. This interpretation is supported by the second set of measurements on CoFeB (high $T_C$, high $M_S$), which exhibit elastically driven precession in only one of the previously determined angular regimes, and amazingly the near complete suppression of precession in the second angular regime. 

We claim that these findings can be reconciled by considering the interaction of elastic waves with the underlying modulated magnetization landscape induced by the spatially periodic heating.  Using the Plane Wave Method (PWM)\cite{Klos16, Klos14}, we calculate the SW eigenfrequencies and corresponding spatial profiles taking into account (de)magnetizing fields as a function of angle between the applied magnetic field and the direction of modulation of magnetization (see Fig.~\ref{fig:experiment}). The calculations are performed for selected modulation depths of the time-dependent magnetization landscape, which are extracted from Two-Temperature Model (TTM) and two-dimensional thermal diffusion considerations\cite{Janusonis2016_1}. The angular dependence of the eigenmode frequencies (at $k=0$) are verified by micromagnetic simulations (MS) and excellent agreement is achieved between the more rigorous and semi-analytical PWM and numeric micromagnetic results. 

\section{Experimental Results}

We begin with a brief recapitulation of the key features of our experimental approach.  As shown in Fig.~\ref{fig:experiment} the experiment relies on the impulsive optical excitation of elastic and magnetic dynamics at the surface of a thin metallic ferromagnetic film by impinging two interfering laser pulses onto its surface.  In our experiments we utilize the second harmonic of the Ti:Sapphire amplified laser as the excitation source, whose  primary action is to (1) impulsively suppress sample magnetization\cite{Beaurepaire1996, Roth2012} in the form of a spatially periodic pattern, and (2) thermoelastically excite acoustic waves that propagate along the surface of the film/substrate heterostructure\cite{Rogers, Crimmins1999}.  We have previously identified the acoustic waves as both Rayleigh Surface Acoustic Waves (SAW) and Surface Skimming Longitudinal Waves (SSLW)\cite{Janusonis2016}, the latter also having been shown recently by Sander el.al.\cite{Sander2017}. The ensemble of excitation processes are then probed by a normally incident probe pulse, and can include time-resolved diffraction of probe light due to the spatial periodicity of strain and/or surface deformation and/or polarization analysis of the transmitted or specularly reflected probe light to extract magnetization dynamics. For magnetization dynamics, we also implement an electromagnet that can rotate around the sample normal.  Further experimental details can be found in Janusonis et.al.\cite{Janusonis2016_1} 

The details of the magnetoelastic interaction depend on the material and substrate combination. However, we can make a few general statements.   For a fixed grating periodicity, the observed frequencies are solely determined by the velocity of acoustic waves in the film/substrate heterostructure, and may depend on the propagation direction for example in a crystalline material.  The amplitude of strain will vary depending on which elastic mode is being driven and the film/substrate thermoelastic properties as well as the grating periodicity.  With regard to magnetization dynamics, there will be an applied field condition wherein the natural precessional frequency of the ferromagnetic resonance or a particular spin wave resonance will match that of the underlying elastic wave, at which point elastic energy will drive precessional motion resonantly via magnetoelastic interactions, provided the spatial symmetries of the particular magnetic and elastic excitations are similar. The resonance condition can be visualized either in the frequency domain as an increase in precessional amplitude in the Fourier transform, or in the time domain by an increase in the temporal range over which precession occurs and/or the amplitude of this precession (i.e. in the maximum polarization rotation in a Faraday geometry).  Furthermore, in the time domain a characteristic phase evolution is observed as the resonance is traversed.  Finally, we mention that there are conditions under which parametric frequency mixing effects have been observed\cite{Chang2017}, wherein magnetization precession is driven at the sum and difference frequencies of the underlying elastic wave(s).  Until now we have speculated that all resonance conditions were the result of the interaction of the elastic waves with the uniform precessional motion, i.e. the FMR.

\begin{figure}
    \centering     
    \includegraphics[width=1\columnwidth]{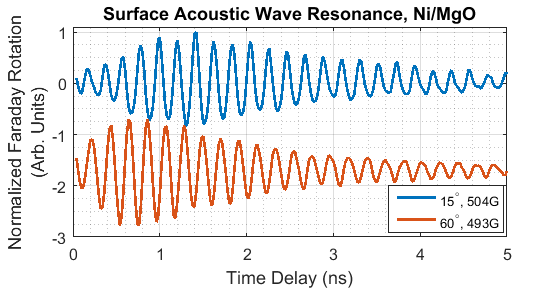}
    \caption{Normalized time-resolved Faraday traces at the Rayleigh Surface Acoustic Wave resonance for Ni/MgO at two representative angles.  The linecuts exhibit opposite phases of precession and differences in their onset times. }
    \label{fig:PhaseflipLineCuts}
\end{figure}

The first indication of non-trivial dynamics in our experiments are contained in Fig. \ref{fig:PhaseflipLineCuts}, where we compare the (normalized) temporal evolution at the resonance condition for two representative magnetic field angles, $15^\circ$ and $60^\circ$ (we will continue to compare these two angles as representative angles for the general features of the experiments).  The data shown here is for the Rayleigh SAW resonance for the Ni/MgO heterostructure at $\approx500$G. Both plots, being on resonance and driven by the same acoustic transient (the elastic frequency is independent of magnetic field angle), precess at the same frequency. However, there are clear differences. First and most notably, the plots show a difference in precessional phase - an unexpected feature for an elastically driven FMR in a uniformly magnetized film. Second, the shapes of the magnetization precession in time are drastically different for the two plots, particularly the onset time at $60^\circ$ is considerably faster than at $15^\circ$.  Both features (delayed onset, opposite precessional phase) are characteristic responses for any Ni/substrate configuration and any type of elastic wave resonance (SAW or SSLW). In the remainder of this report, we provide details of these unexpected results in order to support our picture of elastic excitation of a variety of spin wave modes in the (optically induced) magnetically textured thin film.

\begin{figure}
    \centering    
    \includegraphics[width=1\columnwidth]{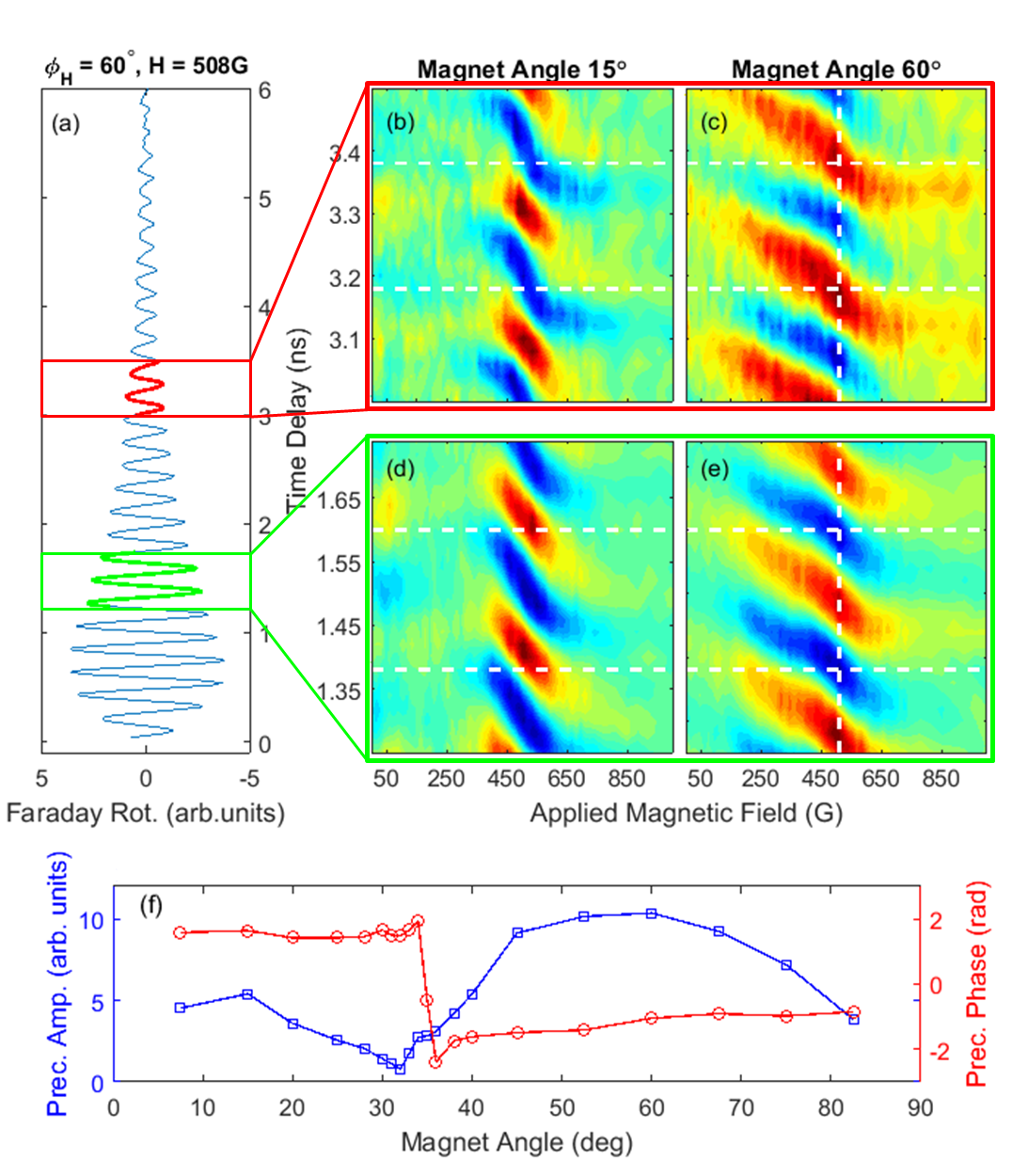}      \caption{ The Surface Acoustic Wave resonance for Ni(40nm)/MgO.  (a) A representative time-resolved Faraday response is taken for magnet angle $\phi=60^\circ$ and at resonance field $H\approx504$G (vertical dashed line in (c, e)). The time response (a) shows the oscillatory dynamics that persist for nearly 6ns. The amplitude of Faraday signal exhibits the resonance dependence on applied field $H$ both for late times (b, c) and early times (d, e). This resonance can be attributed to the interaction between elastic and magnetic degrees of freedom. The backwards S patterns in (b - e) show that the phase changes by $\pi$ as the resonance condition is crossed with increasing field.  A comparison between (b, d) and (c, e) shows an additional change in precessional phase between low and high angles, as indicated by the white horizontal lines (the key finding of this paper). Accompanying the changes in precessional phase, is a suppression of precessional amplitude for intermediate values of magnet angle (f) which shows the precessional amplitude and phase at the peak resonance field.}
    \label{fig:AngleComp}
\end{figure}

In Fig.~\ref{fig:AngleComp}(a), we replicate the time resolved Faraday plot shown in Fig.~\ref{fig:PhaseflipLineCuts} to bring attention specifically to the flipping of the precessional phase for different angular regimes.  A single time trace is extracted from our full field measurements, portions of which are shown in Fig.~\ref{fig:AngleComp}(b-e) for representative magnet angles of $15^\circ$ and $60^\circ$. As indicated in the time trace [Fig.~\ref{fig:AngleComp}(a)] the full field scans are shown for both late time delays (red, panels b-c) and early time delays (green, panels d-e). Each panel (b - e) are individually scaled in amplitude in order to show the shape of the resonance. We draw attention as well to the backward 'S' shape in each of the panels. As mentioned previously this shape reveals that a $\pi$ phase shift occurs as the resonance is traversed and is the hallmark of a driven harmonic oscillator.

This representation brings into focus precessional phase differences between the angles $15^\circ$ and $60^\circ$ as can be seen by following the horizontal lines (compare panels b and c, d and e). In assessing the full angular range, we identify an intermediate angular regime where these two  precessional features interfere resulting in a suppression of measured precessional amplitude. This interference effect is shown in Fig.~\ref{fig:AngleComp}(f) which is extracted from Fourier transforming the time domain data and assessing the precessional amplitude at the peak resonance field.  Accompanying this fitting procedure the phase of precession is also extracted and overlayed in panel (f).  For Ni/MgO only angles between $7.5^\circ$ and $82.5^\circ$ were acquired, but supporting data in the appendix section 1 show full angular dependencies for Ni/Glass heterostructures and null signals at $0^\circ$ and $90^\circ$ as expected for the magnetoelastic interaction. The suppression in amplitude in an angular range around $30^\circ$ separates two excitation 'lobes.' We find that all angles in the first lobe display the phase indicated in Fig.~\ref{fig:AngleComp}(b,d), while all angles in the second lobe display the precessional phase indicated in (c,e) (i.e. the extracted phase in (f) is binary).  The same phase reversal and amplitude suppression phenomena are present for all acoustic frequencies regardless of the acoustic mode (SAW or SSLW), all Ni/substrate combinations (appendix section 1), and is only a function of the relative angle between the excitation wavevector and applied field. The latter is verified by changing the absolute angle of the transient grating excitation (by rotating the phase mask angle) and finding the new magnet angle where suppression occurs.

\begin{figure}[h]
    \centering    
    \includegraphics[width=1\columnwidth]{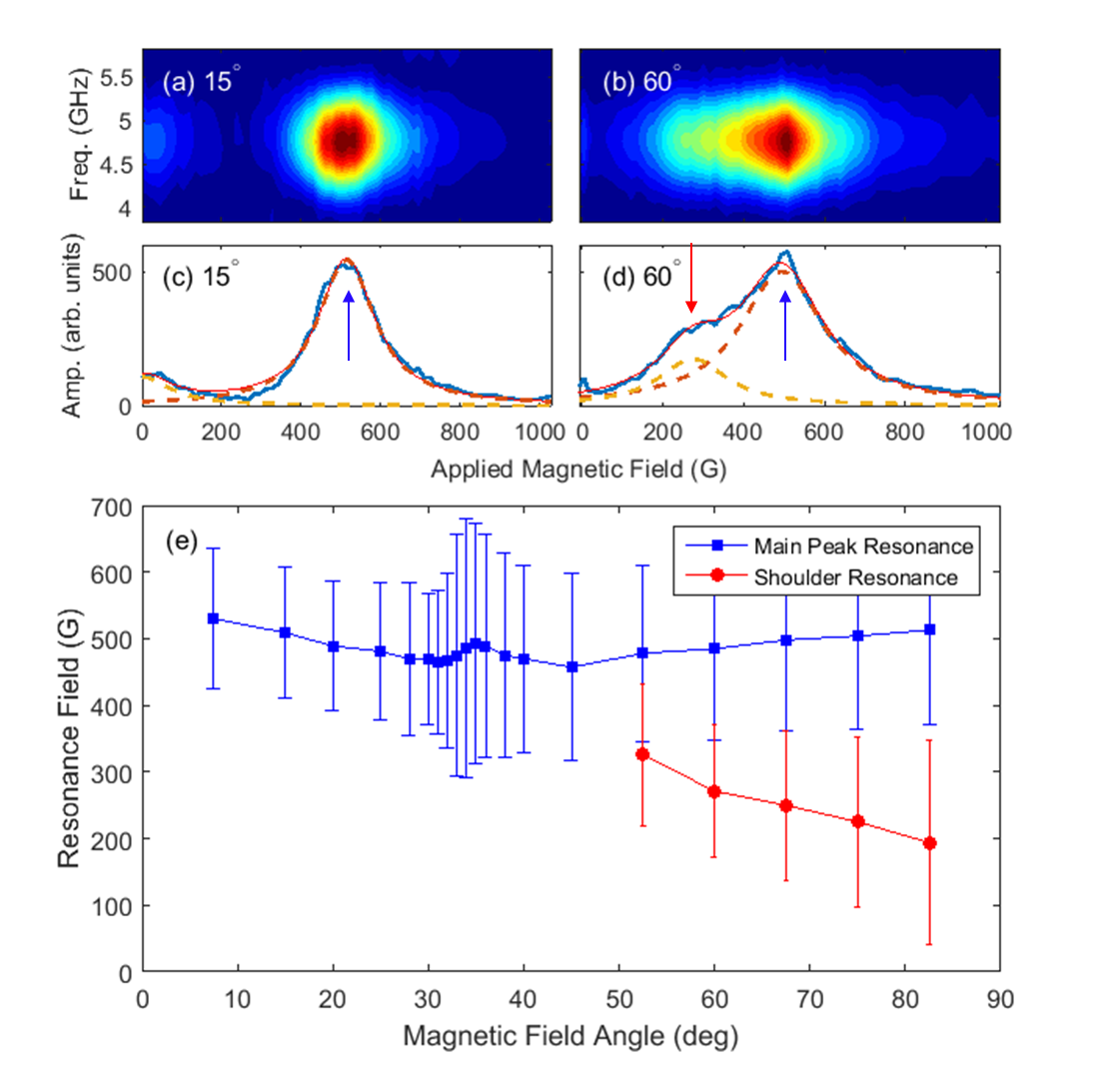}
    \caption{ (a, b) Fourier amplitude of the magnetization precession shows the driving frequency of $\approx 4.8\rm GHz$ which is dictated by the excitation grating period and acoustic velocity. Plots (c, d) present the field dependent lineshapes of (a, b) accompanied by two-Lorentzian fits to the data. At all angles above $40^\circ$, the resonance lineshape is well represented by a two-Lorentzian fit incorporating a main resonance (blue arrow) and a shoulder (red arrow). At all angles below $\approx 20^\circ$ the resonance lineshape is not reproduced by a Lorentzian fit, specifically the low field side exhibits a reduced spectral intensity.  For intermediate angles, the resonance amplitude is small and consequently the lineshape is difficult to fit.  The resonance fields of both the main response and high-angle shoulder are plotted in (e), where the error bars represent the widths of the resonances. The strongly dispersing shoulder is evident, and can be seen explicitly in appendix section 2.}
    \label{fig:PeakShape}
\end{figure} 

It is also clear from the data displayed in Fig.~\ref{fig:AngleComp}(b - e) that the shape of the resonance differs in the two angular regimes.  To extract this behaviour, we apply Fourier transforms to the time resolved data for all  magnetic field angles. Representative Fourier transforms for $15^\circ$ and $60^\circ$ are shown in Fig.~\ref{fig:PeakShape}(a,b) along with linecuts and associated two-Lorentzian fits in panels (c,d).  We make two notes that hold generally for all excitation frequencies and substrate materials: (1) For angles in the second response lobe ($\theta > 40^\circ$), the resonance lineshape is well represented by a two-Lorentzian fit, with a main resonance (blue arrow) and a prominent low-field shoulder (red arrow). The field at which the shoulder resonance occurs strongly reduces as the magnetic field angle is increased. (2) In the low angle regime ($\theta < 20^\circ$) the lineshape is not well represented by Lorentzians, and always exhibits a suppressed spectral weight on the low field side of the resonance.  Intermediate angles (near the suppression) are difficult to assess due to low precessional amplitude and the onset of mixed phase behaviour.

The positions of the main resonance and shoulder are plotted in panel Fig.~\ref{fig:PeakShape}(e) where the error bars represent the widths of the resonances (the widths of the resonances do not change appreciably as a function of angle). Center positions and widths are extracted from the multipeak fitting procedure.  Regardless of the fitting function (Lorentzian or Gaussian) the peak positions are found consistently, while details of the lineshape can only be recovered by utilizing the appropriate function in the appropriate angular range. In appendix section 2 we show the resonance lineshapes as the angle is increased to show the shoulder dispersion.

\begin{figure}
    \centering    
    \includegraphics[width=1\columnwidth]{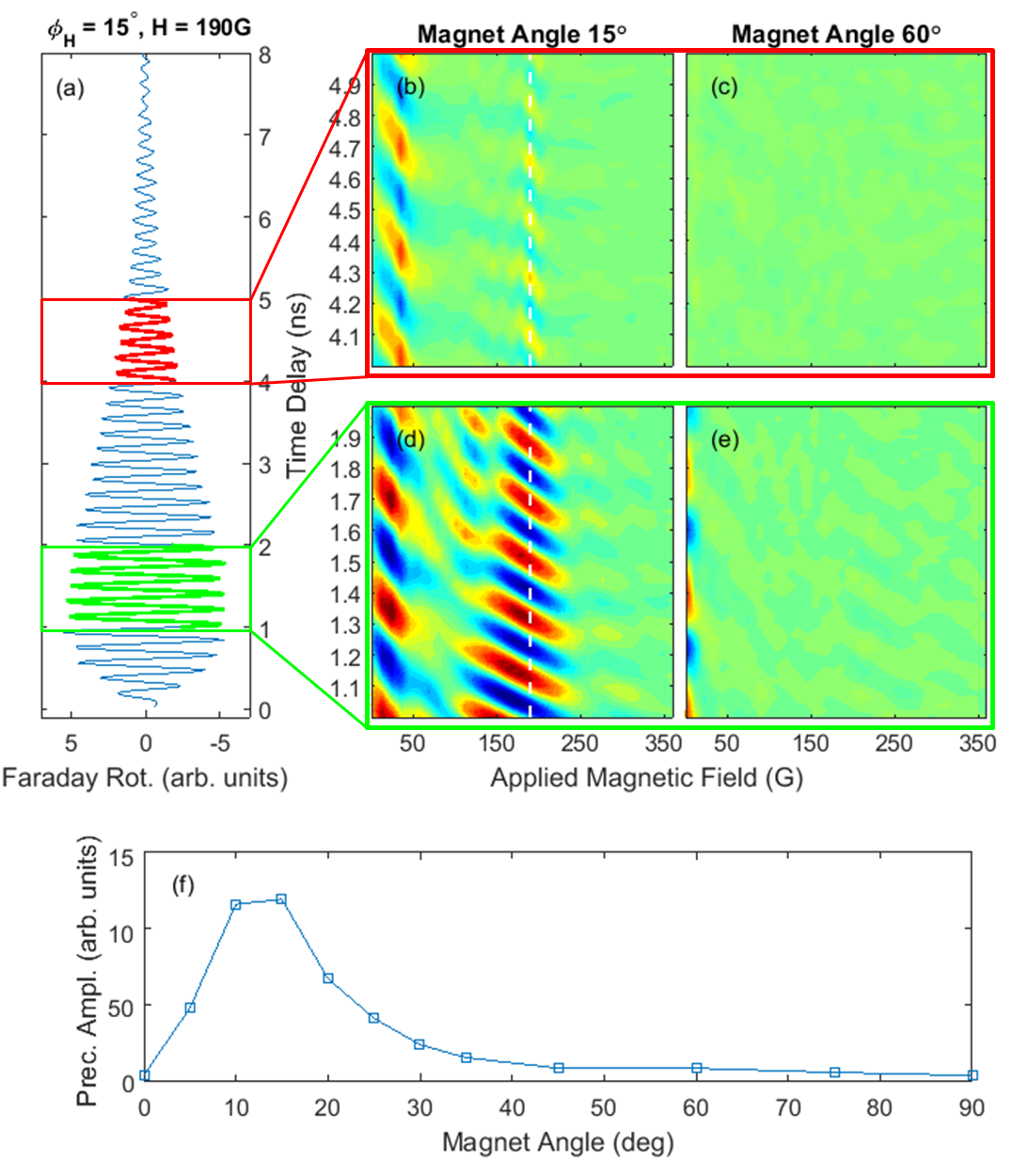}  
    \caption{ The magnetoelastic response for CoFeB(40nm)/glass is observed at low magnet angles only. (a) A representative time-resolved Faraday signal is displayed for the higher elastic resonance (SSLW) at $H = 190$G and $\phi_H=15^\circ$ (see dashed vertical line in (b, d)).   (b - e) CoFeB only exhibits at low magnet angles while the precessional motion at large magnet angles is strongly suppressed (f). Panels (b - e) are plotted on the same scale. In comparing this result to Ni samples, we recognize that only the first precessional lobe is active in CoFeB while two distinct precessional lobes are active in Ni. }
    \label{fig:CoFeB}
\end{figure}

We now turn to the response in the second test material, CoFeB, which shows both similarities and differences in comparison to the Ni sample. First, it is clear that a similar resonance condition can be achieved [Fig.~\ref{fig:CoFeB}(a-f)]. At $15^\circ$ magnet angle we show the response for the SSLW (5.25GHz) resonating at $\approx150$G while the SAW response can be seen at much lower field values ($<50$G) (due to the larger $M_S$, all resonances are downshifted in applied field relative to low $M_S$ materials for a fixed excitation frequency). The precessional motion persists for extended lengths of time while width of the resonance as a function of applied field is narrow, both features related to the low Gilbert damping in this material (precessional damping parameters for these films were previously measured by members of this collaboration\cite{Ulrichs2010} ).  Furthermore, owing to this narrow resonance, nearby resonance features are now apparent.  This is especially the case on the low field side ($\approx$ 100G, 1.8ns) and a suggestion of a resonance on the high field side ($> 250$G).  Both additional resonances clearly have the same frequency as, but appear to be out of phase with, the main SSLW feature. However, perhaps the most striking deviation from the Ni data is the near complete suppression of precessional motion at large angles.  The integrated intensity at the SSLW resonance is shown in panel (f), indicating that this suppression is present over a large angular range. To summarize, a comparison between the two materials indicates that for Ni precession occurs in two distinct angular regimes, accompanied by an intermediate interference, while for CoFeB only the first precessional lobe can be accessed.  In both materials the first precessional lobe peaks at $\approx 15^\circ$. We mention in passing that our films of CoFeB on glass substrates do exhibit a uniaxial in-plane anisotropy (significantly weaker than reported in\cite{Martens2018}), however this is determined not to be the cause of the suppression in high angle response. 
It is our contention that these effects are accounted for if we consider the laterally varying (transient) magnetic texture, its associated spin wave distribution, and finally their resonant interaction with phase locked elastic waves.  We are guided into this line of thinking based on two considerations.  First, a recent paper by Langer et al.\cite{Langer2017} showed both in calculation and experiment that a laterally varying demagnetization landscape (along one dimension) localizes SW distributions in different regions of the MC based on the angle of the applied magnetic field.  Secondly, in comparison to other works in magnetoelastics, ours is the only one in which acoustic waves interact with a spatially modulated magnetization profile, as well as the only one that shows anomalous angular dependence. In experiments most similar to ours, the resonant interaction between surface propagating elastic waves and magnetization\cite{Dreher, Weiler2, Gowtham2015} exhibit coupling behaviour roughly peaked at $45^\circ$, with no indication of a suppression at intermediate angles. These studies occur on uniformly magnetized films,  while the experimental technique (acoustic power transmission) precludes a direct measurement of the magnetization and the details of precessional phase, as we achieve here.

To incorporate the spatial periodicity in our understanding, we take cues from Langer et.al.\cite{Langer2017} and determine the SW eigenmodes in our laterally modulated magnetization profile.  We calculate the temperature profile using the TTM until the electrons and lattice are in thermal equilibrium within the pump excitation volume, and then propagate this temperature gradient in two-dimensions using COMSOL's thermal diffusion capabilities.  The simulation incorporates the thin film and substrate thermal conductivities as well as the thermal boundary resistance between the two dissimilar materials.  Periodic and insulating boundary conditions are used where appropriate.  From this temperature profile a magnetization profile $M_{\rm S}(x)$ is calculated using a Curie-Weiss law for Ni and data measured on similar films of CoFeB\cite{Busse2015}.  For the timescales involved in our experiments (several ns) the temperature and magnetization profile are taken as constant throughout the depth of the film.

\section{The model}

We use two computational methods to simulate the experimental outcomes and to understand the physical mechanism behind them: The Plane Wave Method (PWM) -- based on home made code and Micromagnetic Simulations (MS) -- performed with the aid of ${\rm mumax3}$ package \cite{key-4}. For PWM we used $M_{\rm S}$ calculated numerically using TTM whereas for MS we approximated this profile by a sinusoidal function.  Both methods use the Landau-Lifshitz (LL) equation as an equation of motion:
\begin{equation}
\frac{\partial\bm{M}}{\partial t}=\mu_{0}\gamma\bm{M}\times\bm{H}_{\rm eff},\label{eq:LL}
\end{equation}
where $\mu_{0}$ is permeability of vacuum, $\gamma$ - gyromagnetic ratio,
$\bm{M}$ - magnetization vector. $\bm{H}_{\rm eff}$ denotes the effective
field which is composed of the following terms:
\begin{equation}
\bm{H}_{\rm eff}=\bm{H}_{0}+\bm{H}_{\rm dm}\left(\bm{r},t\right)+\bm{H}_{\rm ex}\left(\bm{r},t\right).\label{eq:field}
\end{equation}
The field $\bm{H}_{0}$ denotes the in-plane applied external magnetic field, $\bm{H}_{\rm dm}\left(\bm{r},t\right)$
is demagnetizing field, and $\bm{H}_{\rm ex}\left(\bm{r},t\right)$ is
exchange field \cite{key-3}. In our system, the external field can be rotated in-plane with respect to the 1D spatial profile of magnetization saturation [Fig.\ref{fig:experiment}]. 
In both PWM and MS calculations we assume the following values of material parameters: $\gamma=176$GHz/T, $\mu_{0}H_{0}=0.05$T (500G), $M_{\rm S,Ni}=0.484\times10^{6}$A/m, exchange length: $\lambda_{\rm ex,Ni}=7.64$nm, period of $M_{\rm S}(x)$: $\Lambda=1.1\mu $m, thickness of Ni layer: $d=40$ nm.  Since the magnetic landscape modulation is smooth, and therefore there are no abrupt changes of static demagnetizing field, the static component of magnetization can be considered as saturated and parallel to the applied field, $\bm{H_0}$. 

\begin{figure}[h]
\centering{}\includegraphics[scale=0.5]{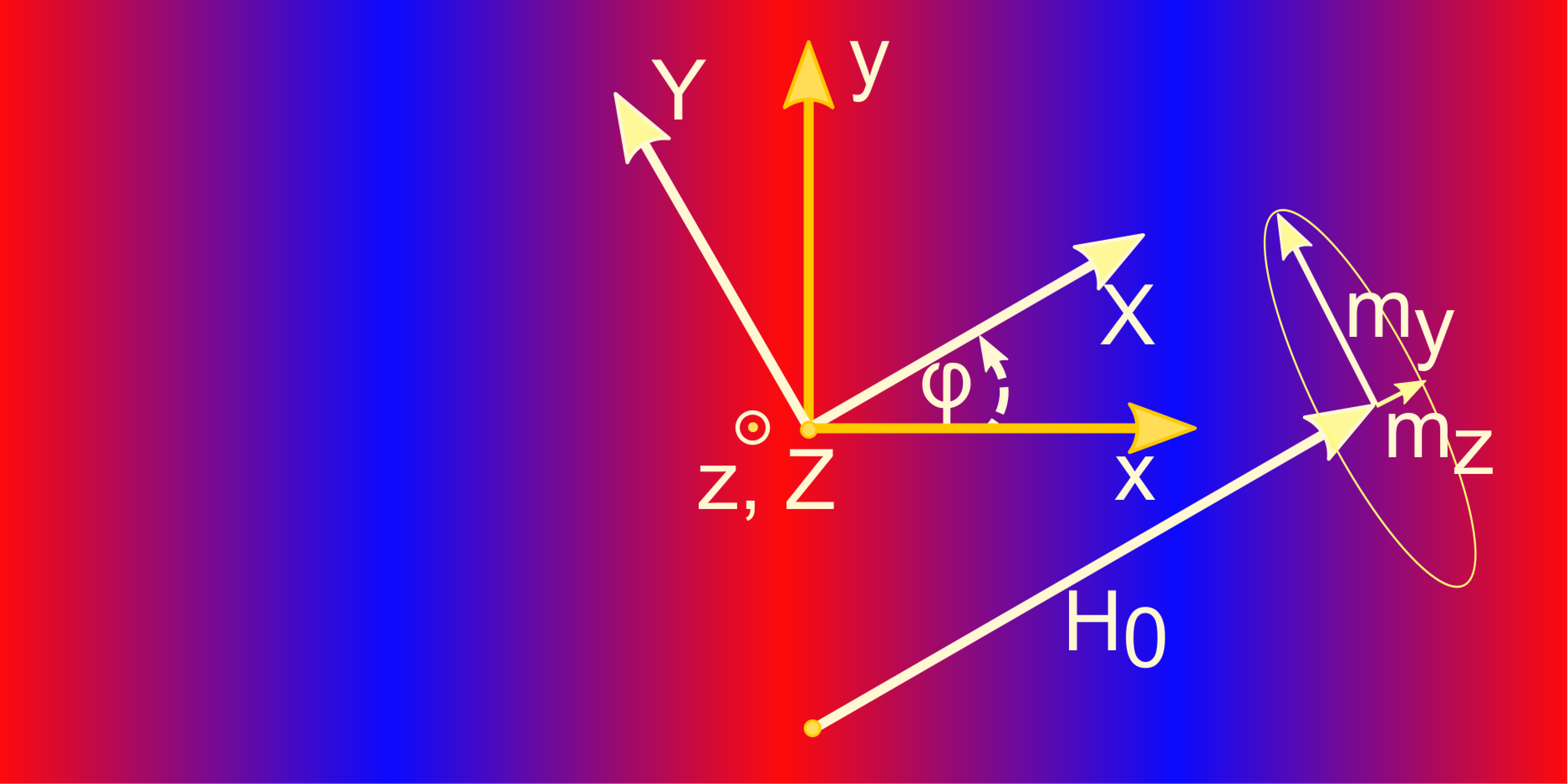} \caption{\label{fig:structure} The top view of the sample. False color representation of the sample temperature (red = hot, blue = cold) for two periods of the MC. The coordinates $\left(x,y,z\right)$ are defined by periodic structure while $\left(x',y',z'\right)$ are oriented with respect to direction of external magnetic field. The in-plane ($y'$) and out-of-plane ($z'$) component of dynamic magnetization depends on $x$ and $y$ coordinates: $m_{\rm y'}(x,y)$, $m_{\rm z'}(x,y)$. }
\end{figure}

The PWM is used, in general, to solve linear differential equations with periodic coefficients where the solutions have a form of Bloch functions. To express the dynamical component of the magnetization $m_{i}$  as well as the dynamic
demagnetizing field $h_{{\rm dm},i}$, we use Bloch functions of the form:
\begin{equation}
\begin{split}
m_{i}\left(\bm{r},\phi,t\right)=m_{i}\left(\bm{r},\phi\right)e^{i\omega t}e^{i\bm{k}\cdot\bm{r}},
\\
h_{{\rm dm},i}\left(\bm{r},\phi,t\right)=h_{{\rm dm},i}\left(\bm{r},\phi\right)e^{i\omega t}e^{i\bm{k}\cdot\bm{r}},\label{eq:blochfun}
\end{split}
\end{equation}
where $i$ denotes in-plane or out-of-plane direction,  $\bm{r}=(x,y,0)$ is in-plane position vector,  $\bm{k}$ is  wave vector and $\omega$ is the angular frequency of SW's precession.   Using the magnetostatic approximation \cite{stanciel}, the demagnetizing fields in a planar magnonic crystal is calculated  analytically \cite{key-5} from Maxwell's equations. For convenience in further calculation, we introduce two coordinate systems: $(xyz)$ -- connected to the periodic landscape and $(x'y'z')$ -- related to the direction of external field [Fig.\ref{fig:structure}]. We consider the $x',y',z'$ components of magnetization $\bm{M}$ and effective field $\bm{H}_{\rm eff}$ upon the spatial coordinates $x,y,z$ and the angle of the applied field  $\phi$\cite{Gallardo18}. The angular dependence of effective field $\bm{H}_{\rm eff}(\phi)$ is included in the model only by the anisotropy of demagnetizing field. Using the method presented by Kaczer\cite{key-5} we can calculate the $x'$ component of static demagnetizing field $\bm{H}_{{\rm dm}}$ :
\begin{equation}
\begin{split}
H_{{\rm dm},x'}\left(\bm{r},\phi\right)&=-\sum_{\bm{G}}M_{S}\left(\bm{G}\right)\cos^{2}\left(\phi\right)\\&\times\left(1-\cosh\left(\left|\bm{G}\right|z\right)e^{-\left|\bm{G}\right|d/4}\right)e^{i\bm{G}\cdot\bm{r}}\label{eq:demag1}
\end{split}
\end{equation}
and $y'$,$z'$ components of the amplitude for dynamic demagnetizing field $\bm{h}_{{\rm dm}}$ for $\bm{k}=0$: 
\begin{equation}
\begin{split}
&h_{{\rm dm},y'}\left(\bm{r},\phi\right)=\\&-\sum_{\bm{G}}\Big[m_{y'}\left(\bm{G}\right)\sin^{2}\left(\phi\right)\left(1-\cosh\left(\left|\bm{G}\right|z\right)e^{-\left|\bm{G}\right|d/4}\right)\\
&-i\;m_{z'}(\bm G)\sin(\phi)\sinh\left(\left|\bm{G}\right|z\right)e^{-\left|\bm{G}\right|d/4}\Big]e^{i\bm{G}\cdot\bm{r}},
\\
&h_{{\rm dm},z'}\left(\bm{r},\phi\right)=\\&-\sum_{\bm{G}}\Big[m_{z'}\left(\bm{G}\right)\cosh\left(\left|\bm{G}\right|z\right)e^{-\left|\bm{G}\right|d/4}\\
&-i\;m_{y'}(\bm G)\sin(\phi)\sinh\left(\left|\bm{G}\right|z\right)e^{-\left|\bm{G}\right|d/4}\Big]e^{i\bm{G}\cdot\bm{r}},
\label{eq:demag2}
\end{split}
\end{equation}
where $\bm{G}=[G_x,0,0]$ is a reciprocal lattice vector. We use $M_{\rm S}\left(\bm{G}\right)$, $m_{z'}\left(\bm{G}\right)$,
$m_{y'}\left(\bm{G}\right)$ to denote the coefficients of Fourier expansions for magnetization saturation $M_{\rm S}\left(\bm{r}\right)$ and periodic factor of dynamical component of magnetization $m_i\left(\bm{r}\right)$. The symbol $d$ stands for thickness of ferromagnetic layer.

The LL equation can be transformed  (in linear approximation)  into the algebraic eigenvalue problem for  eigenvalues (the frequencies of SW eigenmodes) and the eigenvectors (the sets of Fourier components for SW Bloch functions: $m_{i}(\bm{G})$). As a results we obtain the frequency spectrum with the corresponding set of the profiles of dynamical magnetization for SW eigenmodes for selected value of external field angle $\phi$.  

The MS are performed by solving numerically LL equation in real space and
time domain \cite{key-2}. For excitation of the SW precession, we used a microwave external
magnetic field in the form of the {\it sinc} function in the time domain
and spatially homogeneous in the whole sample. After simulating the response for $30$ns, we performed a Fast Fourier Transform of the signal to arrive at the frequency spectra of SW excitation. 

For a given spin wave spatial profile, we consider: (1) the efficiency of detection by Faraday rotation measurements, which relies solely on the spin wave spatial symmetry, and (2) the elastics - to - magnetic excitation efficiency, which relies on both SW and elastic spatial symmetries. We deal with these aspects separately. With respect to detection efficiency, the largest Faraday signal (collected over many spatial periods) will come from any mode whose spatial profile exhibits (relatively) homogeneous phase (an FMR mode\cite{fmr} will possess a large Faraday signal \cite{Hamrle10}, while a mode with odd node number in $\Lambda$ will sum to zero). To assess which of these modes exhibits the largest Faraday detection efficiency, we calculate the net out-of-plane magnetic moment over a period for the modes of $k=0$ by following formula: 
\begin{equation}
I_{n}\propto\frac{\left|\int_{0}^{\Lambda}m_{z}^{n}\left(x\right)dx\right|}{\int_{0}^{\Lambda}\left|m_{z}^{n}\left(x\right)\right|dx},\label{eq:fmr}
\end{equation}
where $m_{z}^{n}\left(x\right)$ is profile of n$^{\rm th}$ eigenmode for out-of-plane component of dynamical magnetization. 

With respect to the excitation cross section, we reiterate that the spatial period of the MC and that of the elastic waves are the same, both being derived from the same optical interference pattern.  Furthermore they are spatially phase locked in that the hot and cold regions of the MC experience the opposite torques (through  magnetoelastic coupling) on each half cycle of the acoustic wave.  Therefore the excitation  efficiency $\sigma$ of SW due to magneto-elastic interaction will depend on spatial profile of SW. To take into account the nonuniform distribution of SW amplitude in one period of MC and the different signs of magnetoelastic torque in hot and cold regions, we integrate the dynamic component of magnetization $m_{z}^{n}\left(x\right)$ with the factor $\cos(2\pi\, x/\Lambda)$ to incorporate the magnetoelastic coupling and opposite sense torque on both half cycles of the elastic wave.:

\begin{equation}
\begin{split}
\sigma\propto\frac{1}{\left(f-f_{0}\right)^{2}+\left({\rm fwhm}/2\right)^{2}}\sin\left(2\phi\right)\\ \times\int_{0}^{\Lambda}\left(m_{ z}^{n}\left(x\right)\cos\left(2\pi\, x/\Lambda\right)\right)dx.\label{eq:sigma}
\end{split}
\end{equation}
The factor $\sin(2\phi)$ reflects the angular dependence of the torque resulting form magneto-elastic interaction of acoustic waves with the magnetization\cite{Dreher} while the Lorentzian factor $\frac{1}{\left(f-f_{0}\right)^{2}+\left({\rm fwhm}/2\right)^{2}}$ reduces significantly the excitation efficiency if the frequency of SW, $f~=~\omega/(2\pi)$, differs from the frequency of the acoustic wave, $f_0$, by more than the elastic bandwidth ($\sim~0.5$GHz).

\section{Numerical simulations}

\begin{figure}[h]
    \centering    
    \includegraphics[width=1\columnwidth]{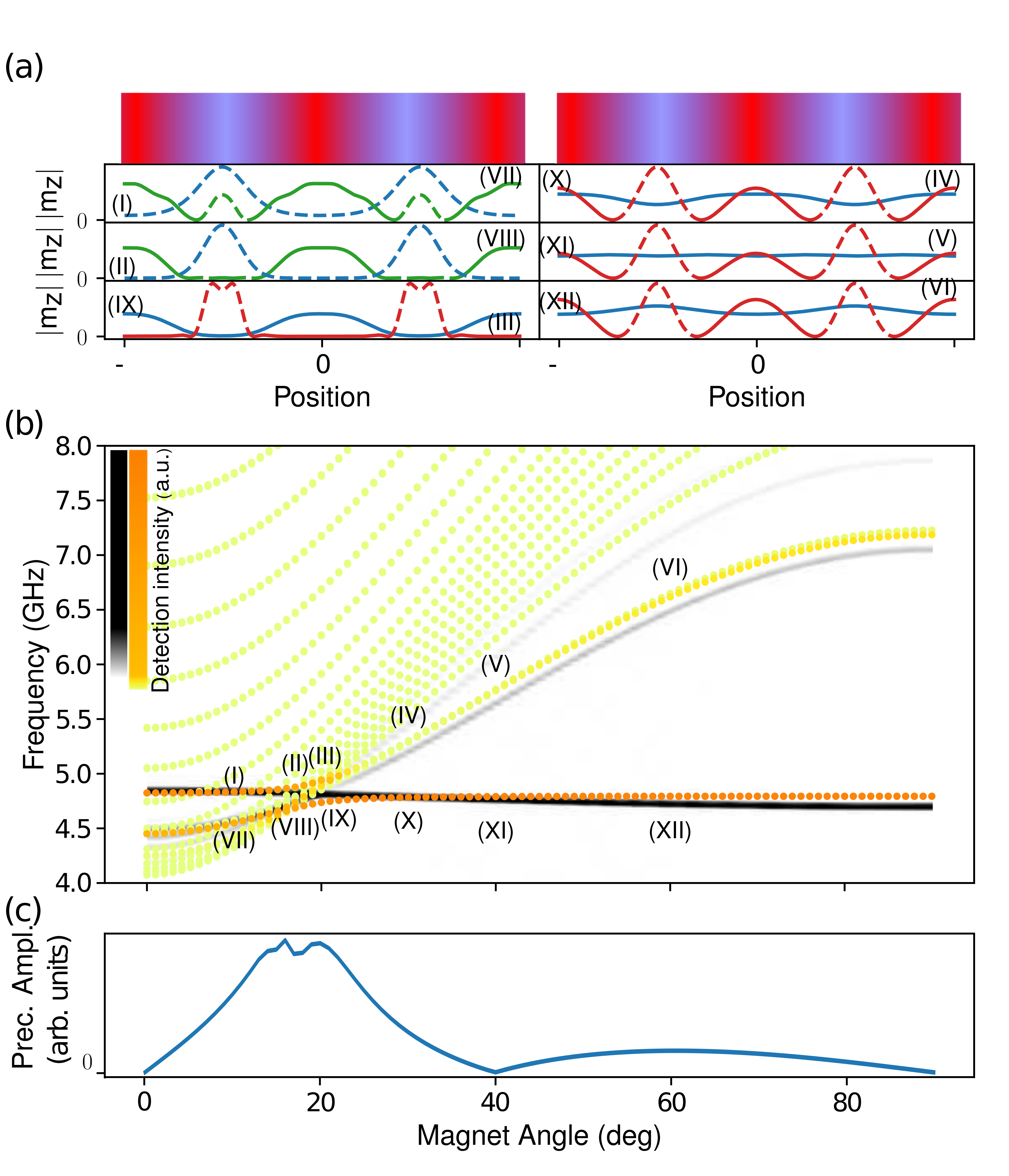}  
    \caption{Plane wave method and micromagnetic calculation for angular dependence of the eigenfrequencies of spin wave modes in a periodically modulated magnetic landscape.  (a) The spatial distribution of the lowest energy  eignemodes (with largest detection efficiency) marked in the main plot by labels: I-XII. The modes were plotted for two periods of the MC. The red (blue) color bars above symolize the hotter (colder) regions of magnonic crystal, respectively.  The sections of solid and dashed lines distinguishes the regions of opposite precession phase. Spatial profiles are color coded to indicate main resonance (blue) and dispersing modes which we associated with the shoulder in \ref{fig:PeakShape}(orange/green). (b)
    Angular dependence of the eigenfrequencies of spin wave modes. The color scales for orange-yellow (PWM) points or black-gray lines (MS) correspond to the simulated detection efficiency (the darker the line/symbol, the larger the value of eq.5). Plane wave method calculation was done for simulated $M_{\rm S}$ profile (from two temperate model), whereas the micromagnetic simulations were performed for sinusoidal approximation of $M_{\rm S}$ profile.  The bottom panel (c) presents the simulated procession amplitude of spin wave pumped by elastic wave inclusive of excitation efficiency (Eq. \ref{eq:fmr}) and detection efficiency (Eq. \ref{eq:sigma}). The outcome is two excitation lobes with a minimum around $40^\circ$.}
    \label{fig:PWM}
\end{figure}

The outcomes of the PWM and MS calculation are provided in Fig.~\ref{fig:PWM}. In Fig.~\ref{fig:PWM}(b) we plot the SW eigenfrequency dependence for all low energy modes as the magnetic field angle is changed from parallel to the periodicity of magnetic landscape ($\phi=0^o$) to the perpendicular direction ($\phi=90^o$) \cite{Gallardo17}. The results of both computational techniques are in agreement for the modes of the largest detection efficiency -- the orange-yellow points (PWM) overlap with black-gray lines (MS). The noticeable features of this result (and all such results for laterally modulated $M_{\rm S}$) is the presence of a nearly constant frequency fundamental mode (characterized by the spatial distribution with zero nodes and homogeneous phase) and the appearance of a network of higher order modes whose frequencies increase as the angle of the magnetic field is increased. At low angles, several modes dip in frequency below the fundamental mode upon anticrossing with fundamental mode at intermediate angle $\phi\approx20^\circ$.   We find that the general shape presented in Fig. \ref{fig:PWM}(b) is reproduced for a large number of modulation depths (i.e. time delays) and is shifted up vertically along the frequency axis as applied field is increased (correspondingly down as the field is reduced).  We did not attempt to perform this calculation for very deep modulations or modulations that deviate strongly from sinusoidal, since they are not relevant to the timescales associated with elastic dynamics.

The spatial profile of any mode can be assessed as a function of magnetic field angle (and strength).  At selected points in the angular dependence, and for the modes showing the largest detection efficiency (orange or dark yellow), we show [in Fig.~\ref{fig:PWM}(a)] the spatial profiles indicated as positions 'I' through 'XII'.  In this case the profiles are only shown for a fixed applied field strength of $500$G, at a fixed time delay (i.e. fixed modulation depth).  In particular, we note that the fundamental mode (I,II,III,X,XI,XII) shows zero nodes, while one of the higher order modes labeled (VII,VIII,IX,IV,V,VI) is displayed with both solid and dashed lines to delineate a change in phase for different portions of the SW profile (i.e. lateral node(s) in the precessional wavefunction).  A symbolic temperature scale is provided above the modal distributions (with blue and red regions corresponding to lower and higher temperature respectively) to indicate where within the lateral dynamic magnetization profile the SW amplitude is maximized.  For example we find a general feature of our calculations to be that at low angles (curves I,II) the mode is strongly localized in the cold regions of the MC which evolves into a uniform profile (curves X,XI,XII) as the angle is increased.  This latter profile we would associate with a true FMR displaying constant phase, and (nearly) constant amplitude over the entire MC.  We can now understand why higher order modes dip in frequency below the fundamental at low angles, since significant SW amplitude is present in the reduced $M_S$, hot regions of the sample (e.g. curve VII), the  frequency of SW eigenmodes is therefore reduced.   Finally, we note that at low angles the mode density is high, while at high angles the modes (at least the lowest two) are energetically well-separated.

The bottom panel of Fig.\ref{fig:PWM} shows a qualitative estimation of the detected signal in a Faraday rotation measurements.  In order to estimate the amplitude of Faraday signal from the spectrum and the profiles of SW eigenmodes we
used the phenomenological formula (\ref{eq:sigma}) and the procedure described in the section 'Model'. The frequency of $f=4.8$GHz driving the SW dynamics and the ${\rm fwhm}=0.5$GHz of this resonance were taken form the experimental outcomes (see Fig.~\ref{fig:PeakShape}). By using this approach we were able to reproduce qualitatively the angular dependence of intensity of Faraday signal. The simulated dependence shown at the bottom of Fig.\ref{fig:PWM} presents two lobes with the distinctive minimum around 30-40 deg. -- similar to those seen in the experiment(see Fig. \ref{fig:AngleComp}). The relative height of two lobes (for lower and higher angles) is different for experimental and numerical outcomes. 

\section{Discussion}
We now discuss the connections between the experimental and calculation results, supporting our claim of the emergence of optically-induced MC and our ability to control the band structure. We begin by considering the behaviour at high magnetic field angles, where the eigenmode solution of the fundamental mode exhibits an FMR-like appearance (SW amplitude is delocalized over the entire MC structure, homogeneous phase profile).  We note that while the detection efficiency of such a mode is large (dark orange in Fig. \ref{fig:PWM}), the excitation efficiency by the elastic waves is actually quite small.  In one period of the MC, the elastic wave has both compressional and dilational phases and thus can not drive the fundamental mode as indicated. However, if we understand the fundamental mode to be driven locally, then at each region of the MC, hot and cold, an FMR can be driven.  Such a locally driven FMR would be exactly the same as in the case of transducer based measurements where a uni-directional multicycle elastic wave drives FMR precessional motion locally and out of phase on each half cycle of the wave\cite{Dreher, Weiler2, Gowtham2015,Thevenard2}.  In our experiments, optically probing the average magnetization precession via Faraday rotation will result in the superposition of FMR responses in the hot and cold regions of the MC which will add out of phase (due to the opposite sign of strain in the two regions).

The picture of a locally driven FMR can explain one of the main observations of our data, namely the difference in high angle response between Ni and CoFeB films.  The marked difference between these two materials is their Curie temperature;  Ni has a low Curie temperature ($600^\circ~K$) while CoFeB has a high Curie temperature ($1300^\circ~K$).  Assuming the optical absorption and thermal diffusion are similar in the two materials (both lustrous metals deposited on similar silicate substrates), this large difference in $T_C$ translates into a smaller magnetization modulation for the case of CoFeB and thus a smaller aggregate Faraday signal.  In fact the case for the locally driven FMR would result in a nearly suppressed high angle response [Fig. \ref{fig:PWM}] which corresponds very well with the results found for CoFeB [Fig. \ref{fig:CoFeB}(f)]. This suggests that for high $M_S$ materials, any spin wave mode with odd spatial symmetry (whether it is a SW with odd spatial symmetry, or a locally driven FMR) would be invisible in a Faraday measurement, while a mode that is strongly localized in one particular region of the MC (i.e. low angles) will be visible in the experiment.  Applying the same considerations to Ni
[Fig. \ref{fig:AngleComp}(f)], we still need to explain the appearance, and the increased signal detected at high angles.  At the points where constructive optical interference occurs, the sample lattice temperature can easily reach $T_C$ and remains at an elevated temperature for several nanoseconds.  Nevertheless, as shown in Fig. \ref{fig:PWM} the modal profile at high angles should maintain its homogeneous phase and amplitude (i.e. the FMR) and thus we would continue to expect a reduced second lobe detection, in effect a result like in CoFeB.  We suspect that the anomalous behaviour in Ni, where in the second lobe we measure a large precessional amplitude, may be the result of a reduction in magnetoelastic coupling strength (not simply the reduction in $M_{\rm S}$) which is prevalent as one approaches the Curie temperature\cite{Lacheisserie1982, Birss1960}. We also mention that similar experiments on arrays of Ni wires, in this case excited by a uniform optical pulse, also excites magnetization and elastic dynamics, as well as the long lived resonant magnetization precession.  In this geometry, the wires themselves should be considered as a proxy for the hot regions of the TG signal (suppressed magnetization, initial dilational strain) and their physical structure such as width, period, material, and substrate were chosen to closely mimic the strain amplitude generated in the TG experiment. Nonetheless the elastically driven magnetization precession up to 10 times smaller than the TG signal for similar excitation fluences when compared at the same magnetic field angle, $15^\circ$ (portions of this work will be published at a future date).  We suspect that this reduced signal level is a signature of hot Ni wires precessing under the action of the elastic waves, but with reduced magnetization and reduced magnetoelastic coupling.  To fully vet this idea, additional fluence dependent measurements would need to be done for a series of materials with varying $M_{\rm S}$ values.

Finally, in the same range of large magnetic field angles one would expect to also measure the higher lying precessional mode (IV - VI) when the field is reduced and this mode crosses the elastic excitation frequency.  Based on the shape of calculated angular dispersion, the larger the angle between the MC wavevector and the magnetic field, the lower in field the resonance will occur.  This is precisely the behaviour that we witness for the high angle shoulder present in the case of Ni.  Again, the visibility of this mode would rely on suppressed detection efficiency in the sample hot region; due both to the reduction in $M_{\rm S}$ as well as the reduced coupling strength. These same arguments indicate that a similar feature would not be present for CoFeB, since the higher lying has an even number of nodes in one period.

At small angles of the magnetic field, the fundamental mode is concentrated in the cold region of the sample regardless of the material in question, and as indicated in Fig.~\ref{fig:PWM} also exhibit a $\pi$ phase flip in the precessional amplitude (plotted as a dotted, rather than solid, line). To assess this absolute phase of precession we follow the fundamental mode profile and precessional phase for sequential small angular steps from 90 to zero degrees (XII - IX, VIII, VII). Assuming that the mode is well behaved and continuously evolves as a function of angle, identifying the precessional properties of this fundamental branch can then be used to reveal the properties of all other modes of precession at low angles. Thus the fundamental mode (XII - IX, I, II) can be identified as precessing with a $\pi$ phase modulation on either side of the anticrossing point.  This feature of the MC spin wave distribution directly relates to the opposite precessional phase for Ni samples in the high and low magnetic field angles.  In both cases, the predominant signal is derived from precession occurring in the cold regions of the sample, but the nature of the MC SW distribution dictates that these two must have opposite phases.

Furthermore, as indicated in the angular dispersion curve [Fig. \ref{fig:PWM}], the MC at low angles of magnetic field exhibits a network of modes at similar energy scales and within the excitation bandwidth of the acoustic wave. These additional modes can be seen explicitly in Fig.~\ref{fig:CoFeB}, where the low Gilbert damping of CoFeB results in narrow field resonances and the appearance of satellite resonance features (i.e. at early time delays [Fig. \ref{fig:PWM}(d)] additional resonances can be seen on both sides of the main precessional mode).  In the high damping case of Ni, the field tuned resonances are wide and individual modes cannot be identified, however the lineshapes of the low angle resonance suggest that more than one mode may be active simultaneously.  For example the out of phase precession of two mode in proximate energy would suppress portions of the observed resonance and distort the lineshape similar to the observed dynamics in Fig.~\ref{fig:PeakShape}(a,c), while simultaneously delaying or slowing the onset of precessional dynamics for example as seen in Fig. \ref{fig:PhaseflipLineCuts}.

\section{Conclusion}

In summary, we have elucidated the magnetoelastic interaction for a range of magnetic field angles relative to the TG excitation wavevector.  The key finding is the identification of distinct angular regimes where precessional motion can be driven elastically.  In the low-$T_C$ Ni sample this is manifested as precessional motion of opposite phase in two angular regimes, along with their interference and suppression of precessional motion at intermediate angles.  For high-$T_C$ CoFeB this is manifested as driven precessional motion in only one of the previously determined angular regimes and the near complete suppression in the other.  To explain these findings we have calculated, using PWM and micromagnetic simulations, the SW amplitude distribution in a laterally (periodically) modulated magnetization profile as a function of modulation depth and magnetic field angle, which in turn has allowed us to infer that in different angular regimes the elastic waves couple to distinct spin wave structures.  At high angles (the second precessional lobe) the elastic wave excites a true FMR response, which we understand to be locally activated at each half period of the elastic wave.  At low angles (the first precessional lobe) we infer that a spin wave mode, localized in the cold region of the sample, is elastically activated.  Connected to these findings we suggest that in low $T_C$ materials such as Ni, one must incorporate an understanding of the temperature dependence of the magnetoelastic constants to understand the observed dynamics, while this is less prevalent in high-$T_C$ materials since even in the hot regions of the sample, optical excitation increases the temperature by only a fraction of $T_C$.
The ability to optically generate a transient magnetic landscape, and control the spatial regions where the localized magnetic groundstates reside, could impact a wide range of opto-magnonics research which currently utilizes artificially textured materials.  

\section*{Acknowledgments}
The  study has  received  financial  support  from  the  National  Science Centre of Poland Grants No. UMO-2012/07/E/ST3/00538 and No. UMO-2016/21/B/ST3/00452 and the EU’s Horizon 2020 Research and Innovation Program under Marie Sklodowska-Curie Grant Agreement No. 644348 (MagIC). RIT gratefully acknowledges support at Los Alamos National by LDRD $\#$20180661ER.  JWK would like to acknowledge the support of the Foundation of Alfried Krupp Kolleg Greifswald. RIT will gladly accommodate reasonable requests for original data sets.  

\vspace{1cm}

\section*{Appendix}
In lieu of a supplementary section, we include here additional data sets displaying the resonance effects in Ni on glass substrates at both SAW and SSLW resonances.

\subsection*{Section 1: Angular dependence for Ni/Glass at SAW and SSLW resonances}
\begin{figure}[h]
    \centering     
    \includegraphics[width=.5\textwidth]{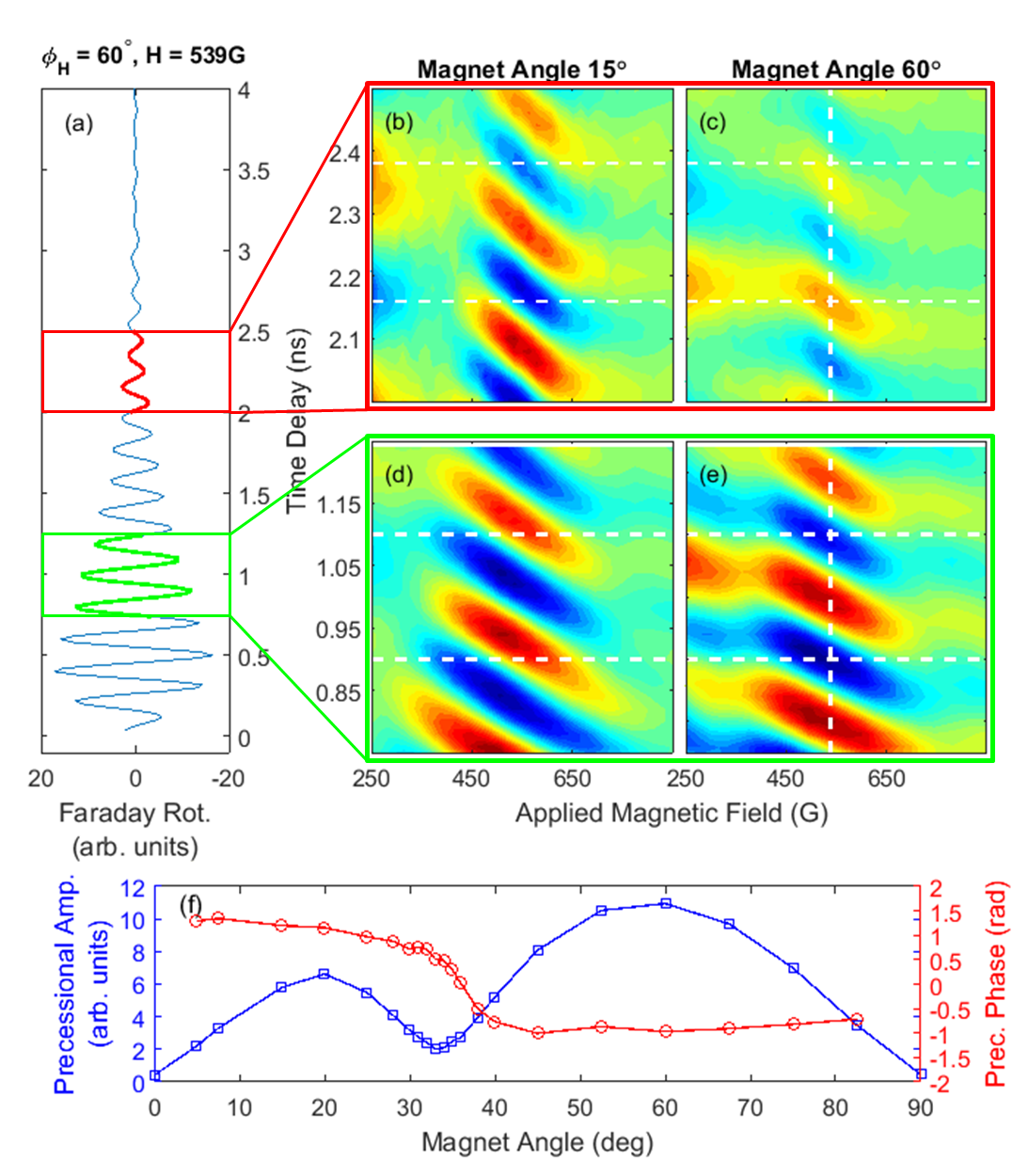} 
    \caption{The surface skimming longitudinal wave on glass substrates ($\Lambda = 1.1\mu$m, $f = 5.15$GHz) shows the same behaviour as the Ni/MgO shown in the main text.   A suppression in precessional amplitude at $\approx 30^\circ$ separates two excitation lobes which opposite precessional phases.}
    \label{fig:SLGPhaseFlip}
\end{figure}

\begin{figure}[h]
    \centering     
    \includegraphics[width=.5\textwidth]{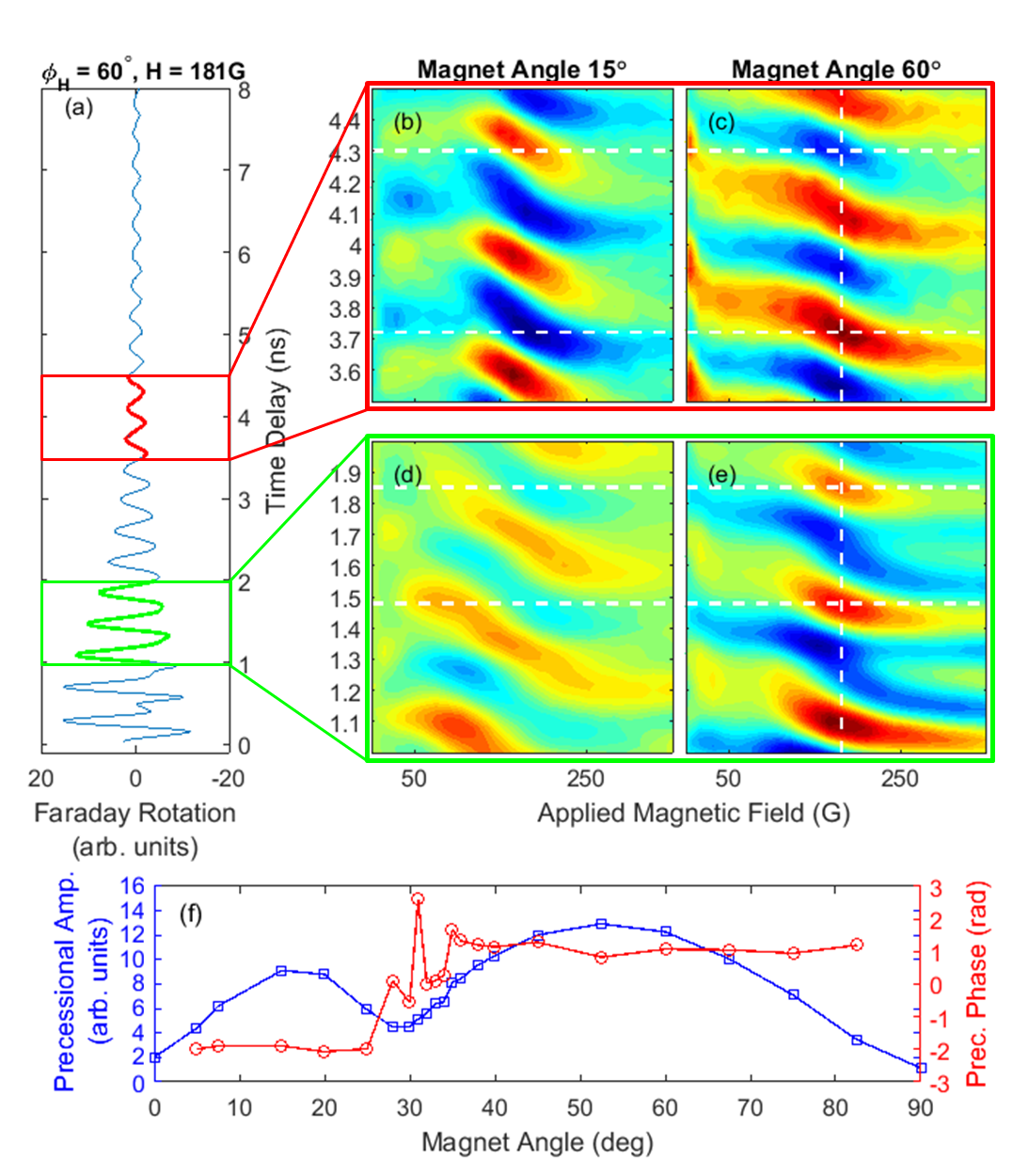}
    \caption{The Rayleigh Surface Acoustic Wave on glass substrates ($\Lambda = 1.1\mu$m, $f = 2.6$GHz) shows the same behaviour as both SSLW on Ni/glass and SAW on Ni/MgO shown in the main text.  A suppression in precessional amplitude at $\approx 30^\circ$ separates two excitation lobes which opposite precessional phases.}
    \label{fig:SLG_SAWPhaseFlip}
\end{figure}

The change in precessional phase and the accompanying intermediate suppression of precessional amplitude is also witnessed for the Ni/Glass heterostructure (here Soda Lime Glass (SLG) or standard microscope slide glass). For this material, a strong but rapidly damped Surface Skimming Longitudinal Wave (SSLW) drives precession at 5.15GHz [Fig. \ref{fig:SLGPhaseFlip}] and the Rayleigh SAW resonance at 2.6GHz [Fig. \ref{fig:SLG_SAWPhaseFlip}].  For both elastic waves, a phase flip and concomitant intensity suppression occurs at roughly $30^\circ$ which is the same angle range that is witnessed in the main text for the Ni/MgO heterostructure. The suppression in amplitude is accompanied by the same binary phase reversal.

\subsection*{Section 2: Dispersion of the Shoulder with Magnetic Field Angle}
In Fig. 4 of the main text we show the resonance field positions of the main and shoulder peaks as a function of angle of the magnetic field.  Here we show all the data sets to further enforce the notion of a strongly dispersing shoulder.  For both glass and MgO substrates we show the integrated Fourier Transform of the resonance responses.  For glass this is the Surface Skimming Longitudinal Wave response at 4.8GHz, while for MgO this is the Rayleigh Surface Acoustic Wave response at 5.15GHz.  The glass data is stronger in amplitude and therefore the signal to noise is better.  However, in both cases we see the main resonance peak change from an asymmetric response at $38^\circ$ to develop a shoulder that disperses to lower field values as indicated by the arrows. The appearance of the shoulder in the Ni samples is thus independent of the type of acoustic wave driving the response, while its observation we suggest in the main text is related to the next SW mode above the fundamental.  The absence of this mode for CoFeB we attribute to the symmetry of the SW and the reduced detection efficiency due to the 
reduced modulation depth in a high $T_C$ material.

\begin{figure}[h]
    \centering     
    \includegraphics[width=.5\textwidth]{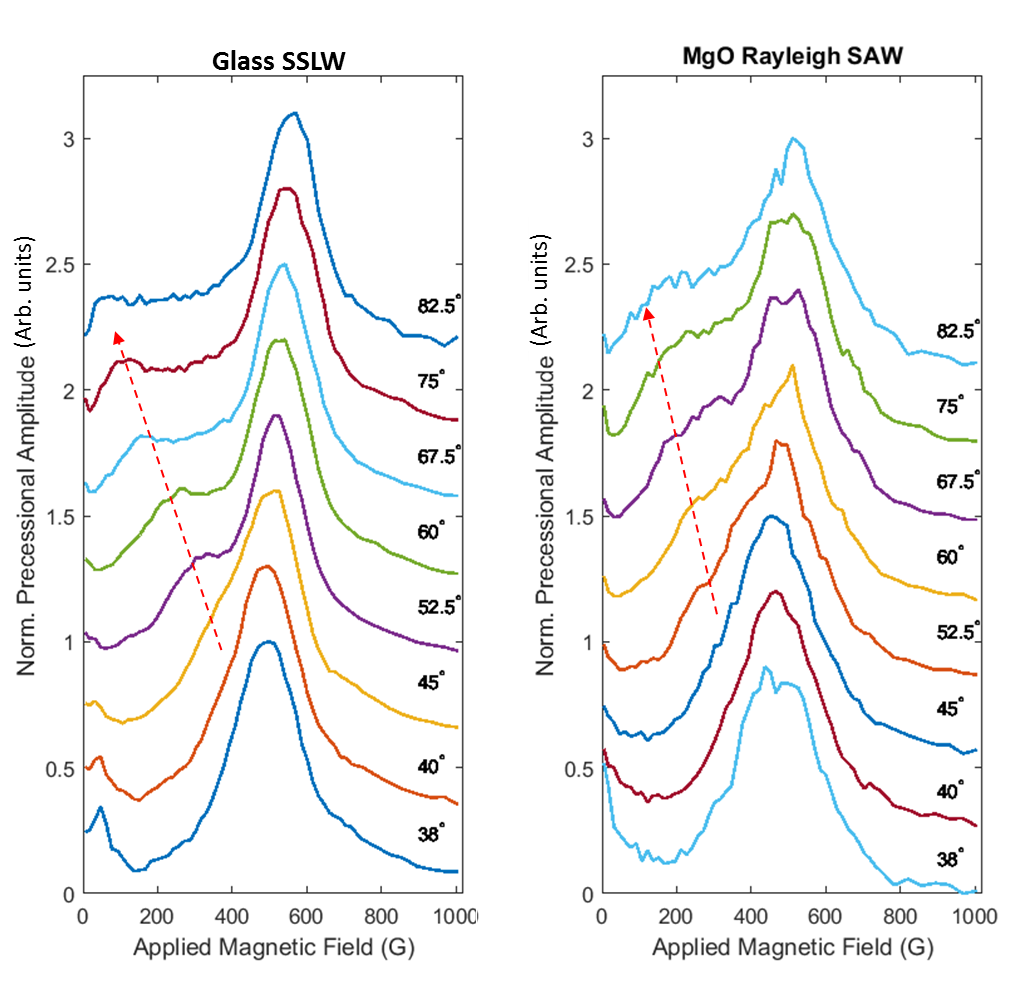} 
    \caption{Resonance response as a function of applied field angle for Ni/Glass (left) and Ni/MgO (right), showing the main resonance and the strongly dispersing shoulder indicated by the arrow.}
    \label{fig:HighAngleShoulderDispersion}
\end{figure}
\vspace{2cm}

\begin{thebibliography}{38}
\bibitem[{pha()}]{phaselocked}
\bibinfo{note}{The excitation geometry we employ ensures that the strain
  profile is phase locked with the magnetic profile. Due to thermoelastic
  processes, the heated regions experience tensile strain in the initial
  oscillation of the elastic wave, while unheated regions experience
  compressive strain. One half period later, the situation is reversed.
  However, the maxima of strain occurring at each half cycle of the elastic
  wave coincide with the maxima/minima of the temperature profile.}
\end{thebibliography}

%

\end{document}